# MINIMIZING COMPRESSION ARTIFACTS FOR HIGH RESOLUTIONS WITH ADAPTIVE QUANTIZATION MATRICES FOR HEVC

*Lee Prangnell and Victor Sanchez*

Department of Computer Science, University of Warwick, Coventry, England, UK

## ABSTRACT

Visual Display Units (VDUs), capable of displaying video data at High Definition (HD) and Ultra HD (UHD) resolutions, are frequently employed in a variety of technological domains. Quantization-induced video compression artifacts, which are usually unnoticeable in low resolution environments, are typically conspicuous on high resolution VDUs and video data. The default quantization matrices (QMs) in HEVC do not take into account specific display resolutions of VDUs or video data to determine the appropriate levels of quantization required to reduce unwanted compression artifacts. Therefore, we propose a novel, adaptive quantization matrix technique for the HEVC standard including Scalable HEVC (SHVC). Our technique, which is based on a refinement of the current QM technique in HEVC, takes into consideration specific display resolutions of the target VDUs in order to minimize compression artifacts. We undertake a thorough evaluation of the proposed technique by utilizing SHVC SHM 9.0 (two-layered bit-stream) and the BD-Rate and SSIM metrics. For the BD-Rate evaluation, the proposed method achieves maximum BD-Rate reductions of 56.5% in the enhancement layer. For the SSIM evaluation, our technique achieves a maximum structural improvement of 0.8660 vs. 0.8538.

## 1. INTRODUCTION

When the scaling list option is enabled in HEVC and its standardized extensions, including SHVC, transform coefficients are quantized according to the weighting values in the QMs. More specifically, after the linear transformation of the residual values, by a finite precision approximation of the Discrete Cosine Transform (DCT), luma and chroma transform coefficients in a Transform Block (TB) are individually quantized according to the weighting integer values in the intra and inter QMs. The default QMs in HEVC are based on the Human Visual System (HVS) and a 2D Contrast Sensitivity Function (CSF) [1-4]. The integer values in the QMs correspond to the quantization weighting of low, medium and high frequency transform coefficients in a TB. Therefore, these QMs possess the capacity to control the quantization step size. A TB contains DC and AC transform coefficients, where the DC transform coefficient is the lowest frequency component and where the AC coefficients correspond to low, medium and high frequency components [3]. Because low frequency transform coefficients are more important for reconstruction, the default QMs apply coarser quantization to medium and high frequency AC transform coefficients. Originally designed for the JPEG standard for still image coding, the QM in [1] is presently employed as the default intra QM in HEVC. This intra QM is derived from a Frequency Weighting Matrix (FWM). The inter QM in HEVC is derived from the intra QM using a linear model [5].

To the best of our knowledge, no previous research has been undertaken in terms of designing a QM technique where the QMs adapt to the display resolution of the target VDU. Therefore, adapting QMs to VDUs is a novel concept. However, alternative QM methods have been proposed to improve upon the default intra and inter QMs in HEVC. For example, the method in [5] involves adjustments to the parameter selection of the HVS-CSF QM technique in [1]. This refinement produces a modified FWM, from which the intra and inter QMs are derived. Although these new parameter insertions may potentially produce coding efficiency improvements, this technique does not take into account the target VDU's display resolution in terms of the quantization of low, medium and high frequency transform coefficients. In [6], the authors propose a novel intra QM method that modifies the weighting values in the QM by employing a normalized exponent variable. Accordingly, the values in the FWM that correspond to medium and high frequency transform coefficients, are modified to decrease the corresponding quantization levels. This results in a quality improvement of the finer details in the images. Similar to the method in [5], this technique does not take into account the target VDU's display resolution.

Based on the resolution of the target VDU, we propose a novel refinement of the HVS-CSF QM method presented in [1]. Compression artifacts are much more visible at high display resolutions, such as HD and UHD, compared with low display resolution [7, 8]. This is true for raw video sequences specifically designed for HD and UHD resolutions and also for those designed for Standard Definition (SD) resolutions that are subsequently coded, decoded and deployed to HD and UHD VDUs. The proposed technique, named *Adaptive Quantization Matrix* (AQM), provides a solution to this problem. In this work, we focus on integrating the proposed intra and inter AQMs into SHVC in order to produce a single two-layered bit-stream, in which each layer is coded to attain the highest possible visual quality for the resolution of the target VDU. At the TB level, similar to the default HEVC QMs, the proposed AQMs also apply different levels of quantization to transform coefficients according to the frequency they represent. The main objective is to decrease the visibility of any compression artifacts that are due to quantization in the decoded layers deployed to high resolution VDUs. More specifically, lower levels of quantization are applied to the ELs, which are then decoded and deployed to high resolution VDUs (e.g., 4K and 8K UHD). Conversely, a higher level of quantization is applied to the BL, which is then decoded and deployed to lower resolution VDUs (e.g., 720p HD and SD).

The rest of the paper is organized as follows. Section 2 provides an overview of the default QMs in HEVC. Section 3 includes detailed expositions of the proposed AQM technique. Section 4 contains an evaluation and discussions of the AQM technique. Finally, Section 5 concludes this paper.

*This work was funded by the Engineering and Physical Sciences Research Council (EPSRC) of the UK. Reference: P83801G.*

## 2. DEFAULT QMs IN HEVC

The HVS-CSF based JPEG 8×8 QM in [1] is employed as the default intra QM in HEVC. HEVC supports up to 32×32 TBs; however, default 16×16 and 32×32 QMs are not present. 16×16 and 32×32 QMs are obtained by upsampling and replicating the 8×8 default intra and inter QMs.

Daly's 2D CSF approach in [2], including the associated Modulation Transfer Function (MTF), is employed to produce a 2D Frequency Weighting Matrix (FWM), $H(f)$, comprising floating point values, from which the default intra QM in HEVC is derived. $H(f)$ is computed in (1):

$$H(f) = \begin{cases} a(b+cf) \times e^{-c(f)^d} & \text{if } f' > f_{max,} \\ 1.0 & \text{otherwise,} \end{cases} \quad (1)$$

where $f$ is the radial frequency in cycles per degree of the visual angle subtended represented in two dimensions such that $f = f(u,v)$. Note that $f'(u,v)$ is the normalized radial spatial frequency in cycles per degree, where $f_{max}$ denotes the frequency of 8 cycles per degree (i.e., the exponential peak). Based on Daly's 2D CSF approach, the MTF is computed with the constant values $a$=2.2, $b$=0.192, $c$=0.114 and $d$=1.1 [1]. Based on rigorous psychophysical empirical testing, constants $a$, $b$, $c$ and $d$ originated in the Mannos-Sarkrison CSF model in [10], and represent regression fits of horizontal and vertical threshold modulation data. After undertaking further psychophysical empirical experimentation, Daly modified the constants $a$ and $b$ of [10] — constants $c$ and $d$ remain unchanged — in order for the CSF to result in higher values for low spatial frequencies. In addition, $f' > f_{max}$ resulting in a value of 1.0 in (1), equates to a lowpass CSF instead of a bandpass CSF, which, with respect to the design of HVS-based QMs, preserves the integrity of the DC component and the low frequency AC components.

In order to account for the fluctuations in the MTF as a function of viewing angle $\theta$, the normalized radial spatial frequency, $f'(u,v)$, is defined using angular dependent function $S(\theta(u,v))$. Both $f'(u,v)$ and $S(\theta(u,v))$ are quantified in (2)-(5).

$$f'(u,v) = \frac{f(u,v)}{S(\theta(u,v))} \quad (2)$$

$$f(u,v) = \frac{\pi}{180 \sin^{-1}(1/\sqrt{1+dis^2})} \times \sqrt{f(u)^2 + f(v)^2} \quad (3)$$

$$S(\theta(u,v)) = \frac{1-s}{2}\cos(4\theta(u,v)) + \frac{1+s}{2} \quad (4)$$

$$\theta(u,v) = \arctan\left(\frac{f(u)}{f(v)}\right) \quad (5)$$

where $dis$ represents the perceptual viewing distance of 512mm and $s$ is the symmetry parameter with a value of 0.7 [10]. Parameter $s$ ensures that the floating point values in $H(f)$ are symmetric. As $s$ decreases, $S(\theta(u,v))$ decreases at approximately 45°; this, in turn, increases $f'(u,v)$ and decreases $H(f)$. The discrete horizontal and vertical frequencies are computed in (6):

$$f(u) = \frac{u-1}{\Delta \times 2N}, \text{ for } u = 1, 2..., N;$$
$$f(v) = \frac{v-1}{\Delta \times 2N}, \text{ for } v = 1, 2..., N; \quad (6)$$

where $\Delta$ denotes the dot pitch value of 0.25mm (approximately 100 DPI) and $N$ is the number of horizontal and vertical radial spatial frequencies. A static dot pitch value of 0.25mm is utilized to compute FWM $H(f)$, which is shown in (7).

$$H = \begin{pmatrix} 1.0000 & 1.0000 & 1.0000 & 1.0000 & 0.9599 & 0.8746 & 0.7684 & 0.6571 \\ 1.0000 & 1.0000 & 1.0000 & 1.0000 & 0.9283 & 0.8404 & 0.7371 & 0.6306 \\ 1.0000 & 1.0000 & 0.9571 & 0.8898 & 0.8192 & 0.7371 & 0.6471 & 0.5558 \\ 1.0000 & 1.0000 & 0.8898 & 0.7617 & 0.6669 & 0.5912 & 0.5196 & 0.4495 \\ 0.9599 & 0.9283 & 0.8192 & 0.6669 & 0.5419 & 0.4564 & 0.3930 & 0.3393 \\ 0.8746 & 0.8404 & 0.7371 & 0.5912 & 0.4564 & 0.3598 & 0.2948 & 0.2480 \\ 0.7684 & 0.7371 & 0.6471 & 0.5196 & 0.3930 & 0.2948 & 0.2278 & 0.1828 \\ 0.6571 & 0.6306 & 0.5558 & 0.4495 & 0.3393 & 0.2480 & 0.1828 & 0.1391 \end{pmatrix} \quad (7)$$

Note that, although the display resolution is already accounted for with the dot pitch value in (6) (because the cathetus pixel resolution values are required to compute the pixel density aspect of the dot pitch computation), the dot pitch can be the same value for a multitude of VDU resolutions depending on the pixel density of the VDU.

The normalized values in $H(f)$ highlight the visually perceptual importance of transform coefficients in the frequency domain. These normalized values are then rounded to integer values to create the default 8×8 intra QM, $QM_{intra}$, by utilizing a scaling value of 16 [11, 1]. The resulting $QM_{intra}$, as quantified in (8), is used to compute the default 8×8 inter QM by using a simple linear model [12, 5].

$$QM_{intra} = \left[\frac{16}{H}\right] = \begin{pmatrix} 16 & 16 & 16 & 16 & 17 & 18 & 21 & 24 \\ 16 & 16 & 16 & 16 & 17 & 19 & 22 & 25 \\ 16 & 16 & 17 & 18 & 20 & 22 & 25 & 29 \\ 16 & 16 & 18 & 21 & 24 & 27 & 31 & 36 \\ 17 & 17 & 20 & 24 & 30 & 35 & 41 & 47 \\ 18 & 19 & 22 & 27 & 35 & 44 & 54 & 65 \\ 21 & 22 & 25 & 31 & 41 & 54 & 70 & 88 \\ 24 & 25 & 29 & 36 & 47 & 65 & 88 & 115 \end{pmatrix} \quad (8)$$

## 3. PROPOSED ADAPTIVE QM TECHNIQUE

Our technique is based on parameter $A_{i,j}$, which is applied to each element of $H(f)$ located at position $(i,j)$, denoted as $H_{i,j}$. $A_{i,j}$ allows modifying $H_{i,j}$ according to the TB size and the VDU's resolution in order to produce an adaptive 2D FWM $H'(f)$. The element of $H'(f)$ located at position $(i,j)$, denoted as $H'_{i,j}$, is computed in (9).

$$H'_{i,j} = H_{i,j}^{A_{i,j}} \quad (9)$$

The computations derived from equations (2) to (6) still apply to $H'(f)$. Parameter $A_{i,j}$ is calculated by an exponential function that uses as input the Euclidean distance between two coefficient locations in a TB and also a display resolution value. This exponential function results in large values corresponding to low frequency transform coefficients, and small values corresponding to high frequency transform coefficients. Parameter $A_{i,j}$ is computed in (10):

$$A_{i,j}(d_{i,j}, w) = e^{-\left(\frac{d_{i,j}}{w}\right)} \in [0,1] \quad (10)$$

where $d_{i,j}$ is the normalized Euclidean distance between the DC transform coefficient and the current coefficient located at position $(i,j)$ in a TB, and $w$ is the display resolution parameter. Euclidean distance $d_{i,j}$ is computed in (11):

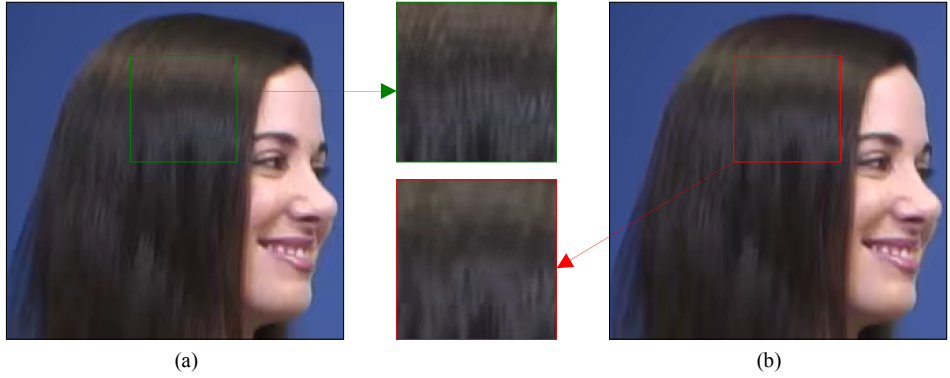

Fig 1. A single frame of the 720p HD KristenAndSara video sequence, coded using a QP = 30, which is the default QP in HEVC HM 16.6 and SHVC SHM 9.0 [11]. Fig 1 (a) shows the improvement of the reconstruction quality of the frame using the AQMs designed for a 4K VDU with QP = 30 versus the default QMs, as shown shown in Fig 1 (b). Note how the medium and high frequency details are preserved in Fig 1 (a). The blurring compression artifacts, caused by quantization, are much more noticeable in Fig 1 (b).

$$d_{i,j} = \sqrt{\frac{(i_1 - i_2)^2 + (j_1 - j_2)^2}{(i_1 - i_{max})^2 + (j_1 - j_{max})^2}} \in [0,1] \quad (11)$$

where $(i_1, j_1)$, $(i_2, j_2)$, $(i_{max}, j_{max})$ represent the position of the floating point values in $H(f)$ associated with the DC coefficient, the current coefficient and the farthest AC coefficient, respectively. For example, for an 8×8 TB the floating point value associated with the DC transform coefficient is located at position ($i=0$, $j=0$) and the farthest AC coefficient is located at position ($i=7$, $j=7$). Each $A_{i,j}$ value decreases as the display resolution parameter $w$ decreases. Note that $w$ is a key parameter to compute the values in $H'(f)$ according to the resolution of the target VDU. The $w$ parameter, quantified in (12), focuses on a specific display resolution:

$$w = h_t^{-p} \in (0,1] \quad (12)$$

where $h_t$ is the VDU's theoretical maximum hypotenuse value, in pixels, and $p$ is the VDU's normalized hypotenuse value, in pixels; $p$ is computed in (13) and $h_t$ is computed in (15):

$$p = \frac{h_a}{h_t} \in (0,1] \quad (13)$$

where $h_a$ is the VDU's actual hypotenuse value in the pixel domain, which is calculated in (14):

$$h_a = \sqrt{x^2 + y^2} \quad (14)$$

$$h_t = \sqrt{x_{max}^2 + y_{max}^2} \quad (15)$$

where $(x, y)$ represent the horizontal and vertical dimensions of the target VDU, respectively, and $(x_{max}, y_{max})$ represent, respectively, the maximum possible horizontal and vertical dimensions of the target VDU.

Parameter $w$ rapidly decreases as $p$ increases (see Fig. 2), which in turn results in higher values in $H'(f)$. Consequently, lower weighting integer values, corresponding to the low, medium and high frequency AC coefficients, are derived for the intra and inter AQMs.

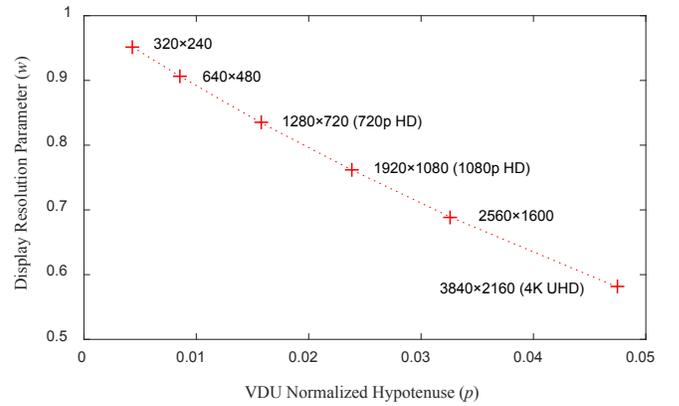

Fig. 2. Display resolution parameter $w$ rapidly decreases as the VDU's normalized hypotenuse value $p$, and its display resolution, increases.

Because the default 8×8 intra QM, $QM_{intra}$, in [1] is designed for the JPEG standard, we base the theoretical maximum pixel values, $x_{max}$ and $y_{max}$, on the maximum possible image size, in pixels, permitted in the JPEG standard. Therefore, $x_{max}=65535$ and $y_{max}=65535$ [14].

The 8×8 matrix in (16) is the derived intra AQM, $AQM_{intra}$, as computed using (10)-(15) for a target VDU of 3840×2160 pixels (4K). The same linear model used to generate the default inter QM from $QM_{intra}$, as specified in [12, 5], is utilized to create the inter AQMs.

$$AQM_{intra} = \left[\frac{16}{H'}\right] = \begin{pmatrix} 16 & 16 & 16 & 16 & 16 & 17 & 18 & 18 \\ 16 & 16 & 16 & 16 & 17 & 17 & 18 & 18 \\ 16 & 16 & 16 & 17 & 18 & 18 & 18 & 19 \\ 16 & 16 & 17 & 18 & 19 & 19 & 20 & 20 \\ 16 & 17 & 18 & 19 & 20 & 21 & 21 & 21 \\ 17 & 17 & 18 & 19 & 21 & 22 & 22 & 22 \\ 18 & 18 & 18 & 20 & 21 & 22 & 22 & 23 \\ 18 & 18 & 19 & 20 & 21 & 22 & 23 & 23 \end{pmatrix} \quad (16)$$

In order to signal the proposed AQMs to the SHVC decoder, we use a layer specific scaling list option, which allows several sets of custom QMs to be signaled to the decoder for multiple layers; i.e., a different set of QMs is signaled for each layer. We also exploit the 8×8 QM upsampling and replication process for the proposed AQM technique.

Table 1. BL and EL average BD-Rate results of the proposed AQM technique compared with anchors. The results in green indicate performance improvements, the results in black indicate no improvements and the results in red indicate negative results.

| Proposed AQM Technique versus Default QMs | | | | | | | | | | Proposed AQM Technique versus Sony QMs | | | | | | | | | |
|---|---|---|---|---|---|---|---|---|---|---|---|---|---|---|---|---|---|---|---|
| **Class** | **All Intra** | | | **Low Delay B** | | | **Random Access** | | | **Class** | **All Intra** | | | **Low Delay B** | | | **Random Access** | | |
| | Y % | U % | V % | Y % | U % | V % | Y % | U % | V % | | Y % | U % | V % | Y % | U % | V % | Y % | U % | V % |
| A (BL) | -0.6 | -0.5 | -0.3 | -2.3 | -2.6 | -2.6 | -3.0 | -5.0 | -5.4 | A (BL) | -2.2 | 0.0 | -0.2 | -3.2 | -3.9 | -4.7 | -4.8 | -6.9 | -7.2 |
| B (BL) | -0.4 | -0.1 | 0.0 | -2.1 | -2.2 | -2.5 | -2.9 | -4.5 | -5.0 | B (BL) | -1.1 | -1.1 | -1.7 | -2.3 | -4.7 | -6.5 | -3.0 | -6.6 | -8.2 |
| C (BL) | -0.4 | 0.2 | 0.2 | -3.1 | -3.3 | -2.9 | -3.6 | -5.4 | -5.0 | C (BL) | -2.3 | -3.9 | -4.8 | -1.9 | -5.5 | -5.3 | -4.8 | -11.2 | -10.8 |
| D (BL) | -0.4 | 0.2 | 0.2 | -2.1 | -2.6 | -2.6 | -2.3 | -3.9 | -2.9 | D (BL) | -2.3 | -0.1 | -3.2 | -1.8 | -2.1 | -3.0 | -2.7 | -5.4 | -5.1 |
| E (BL) | -0.2 | 0.2 | 0.3 | -2.9 | -3.6 | -4.4 | -3.3 | -2.8 | -2.8 | E (BL) | -0.8 | 0.2 | 0.5 | -3.5 | -4.7 | -5.7 | -4.3 | -3.4 | -3.1 |
| **Average** | -0.4 | 0.0 | 0.1 | -2.5 | -2.8 | -3.0 | -3.0 | -4.3 | -4.2 | **Average** | -1.7 | -1.0 | -1.9 | -2.5 | -4.2 | -5.0 | -3.9 | -6.7 | -6.9 |
| **Class** | **All Intra** | | | **Low Delay B** | | | **Random Access** | | | **Class** | **All Intra** | | | **Low Delay B** | | | **Random Access** | | |
| | Y % | U % | V % | Y % | U % | V % | Y % | U % | V % | | Y % | U % | V % | Y % | U % | V % | Y % | U % | V % |
| A (EL) | -0.8 | -0.5 | -0.1 | -19.7 | -19.1 | -19.0 | -37.4 | -39.7 | -40.1 | A (EL) | -2.3 | 0.0 | -0.1 | -52.7 | -56.0 | -56.6 | -56.5 | -58.7 | -59.2 |
| B (EL) | -0.4 | -0.1 | 0.0 | -12.3 | -14.2 | -14.4 | -40.4 | -43.7 | -44.5 | B (EL) | -1.3 | -1.1 | -1.6 | -36.2 | -42.9 | -43.0 | -50.6 | -55.5 | -56.9 |
| C (EL) | 19.0 | 21.1 | 21.5 | -2.9 | -2.6 | -2.2 | -30.0 | -32.0 | -31.9 | C (EL) | -45.0 | -47.8 | -48.2 | -19.9 | -24.1 | -23.7 | -44.4 | -49.1 | -49.1 |
| D (EL) | -3.2 | -3.7 | -3.6 | -6.3 | -6.9 | -7.0 | -32.5 | -35.2 | -34.5 | D (EL) | -52.6 | -56.9 | -57.2 | -28.5 | -31.6 | -31.8 | -39.1 | -42.5 | -42.3 |
| E (EL) | -1.6 | -2.0 | -2.0 | -6.0 | -6.7 | -6.9 | -29.3 | -31.3 | -31.1 | E (EL) | -33.5 | -35.4 | -35.0 | -6.6 | -6.9 | -7.1 | -33.5 | -35.4 | -35.0 |
| **Average** | 2.60 | 2.96 | 3.16 | -9.44 | -9.90 | -9.90 | -33.92 | -36.38 | -36.42 | **Average** | -26.94 | -28.24 | -28.42 | -28.78 | -32.30 | -32.44 | -44.82 | -48.24 | -48.5 |

Table 2. EL average luma SSIM results of the proposed AQM technique (QP = 37) compared with anchors (QP = 37). The results in green indicate performance improvements and the results in red indicate negative results. The results in grey text indicate the results of the anchors.

| Proposed AQM Technique versus Default QMs | | | | Proposed AQM Technique versus Sony QMs | | | |
|---|---|---|---|---|---|---|---|
| **Class** | **All Intra** | **Low Delay B** | **Random Access** | **Class** | **All Intra** | **Low Delay B** | **Random Access** |
| A (EL) | 0.9055 vs. 0.9110 | 0.9157 vs. 0.9110 | 0.9267 vs. 0.9185 | A (EL) | 0.9055 vs. 0.9043 | 0.9157 vs. 0.9099 | 0.9267 vs. 0.9176 |
| B (EL) | 0.9321 vs. 0.9317 | 0.9277 vs. 0.9250 | 0.9403 vs. 0.9343 | B (EL) | 0.9321 vs. 0.9318 | 0.9277 vs. 0.9246 | 0.9403 vs. 0.9339 |
| C (EL) | 0.8429 vs. 0.8426 | 0.8396 vs. 0.8379 | 0.8545 vs. 0.8506 | C (EL) | 0.8429 vs. 0.8426 | 0.8396 vs. 0.8379 | 0.8545 vs. 0.8506 |
| D (EL) | 0.8543 vs. 0.8513 | 0.8391 vs. 0.8353 | 0.8660 vs. 0.8538 | D (EL) | 0.8543 vs. 0.8516 | 0.8391 vs. 0.8339 | 0.8660 vs. 0.8538 |
| E (EL) | 0.9397 vs. 0.9402 | 0.9330 vs. 0.9316 | 0.9437 vs. 0.9405 | E (EL) | 0.9397 vs. 0.9398 | 0.9330 vs. 0.9310 | 0.9437 vs. 0.9402 |

## 4. PERFORMANCE EVALUATION

We undertake thorough evaluations to ascertain the efficacy of the AQM technique on SHVC using the BD-Rate [15, 16, 17] and SSIM [18] metrics. Two-layered bit-streams are created, in which higher levels of quantization are applied to the BL and lower levels of quantization are applied to the EL. For each layer, we employ the proposed AQMs. The BL is aimed at HD 720p VDUs (1280×720), while the EL is aimed at 4K VDUs (3840×2160).

For each evaluation, the AQM technique is tested using the All Intra (Main), Low Delay (Main) and Random Access (Main) configurations [19, 20]. For the BD-Rate evaluation, the QPs used are 22, 27, 32, 37 for BL and EL. For the SSIM evaluation, the luma component of each decoded sequence encoded using QP = 37 is tested. For both evaluations, the test sequences utilized are: Traffic (Class A - 2560×1600), Cactus (Class B - 1920×1080), BasketballDrill (Class C - 832×480), BasketballPass (Class D - 416×240) and FourPeople (Class E - 1280×720). In Table 1 and Table 2, we tabulate the average BD-Rate (for EL and BL) and SSIM performance improvements, respectively, of the proposed AQM technique compared with the default QMs in addition to a state-of-the-art HVS QM technique, developed for HEVC by Sony (anchors) [5].

In the BL versus BL tests the most significant average BD-Rate reductions attained by our method, compared with the default QMs in HEVC, are as follows: 3.6% (Y), 5.4% (Cb) and 5.0% (Cr) for the Class C sequence using the Random Access (Main) configuration. In comparison with the QM technique developed by Sony, the most noteworthy average luma and chroma BD-Rate improvements achieved by our method are as follows: 4.8% (Y), 11.2% (Cb) and 10.8% (Cr) using the Random Access (Main) configuration (see Table 1).

In the EL versus EL tests, compared with the default QMs in SHM, the largest improvement is recorded for the Class B HD sequence using the Random Access (Main) configuration. Specifically, the following BD-Rate improvements are achieved: 40.4% (Y), 43.7% (Cb) and 44.5% (Cr). In contrast with the QM technique from Sony, considerable BD-Rate improvements are achieved on the Class A (UHD 4K) sequence using the Random Access (Main) configuration, which are as follows: 56.5% (Y), 58.7% (Cb) and 59.2% (Cr).

For the BD-Rate evaluation, our technique performs well using the Random Access (Main) configuration because of its temporal coding structure. The proposed AQM technique attains the best performance when there is a larger group of B pictures in the GOP structure. The very high EL versus EL BD-Rate improvements are mainly due to the increased accuracy of inter-layer prediction for the EL. More specifically, in the BL our AQMs contain lower weighting integer values in comparison with the default QM technique and the Sony QM technique (anchors). Consequently, this results in improved reconstruction of the BL and, thus, allows for a more accurate prediction for the EL [21] (see the results in Table 1).

The SSIM evaluation reveals significant reconstruction quality improvements of the proposed AQM technique compared with anchors, with a maximum SSIM value difference of 0.8660 vs. 0.8538 for the Class D sequence using the Random Access (Main) configuration (see Table 2). This represents the preservation of important detail in the reconstructed video data (see Fig. 1).

Compared with the default QMs and the Sony QMs (anchors), the proposed method yields average encoding time reductions of 0.75% and 1.19%, respectively. In addition, our technique yields average decoding time reductions of 4.67% and 2.82%, respectively. A more accurate prediction of the EL from the BL decreases the workload of the entropy coding process [21]. Therefore, this reduced workload results in encoding and decoding time improvements.

## 5. CONCLUSIONS

A novel AQM technique for HEVC is proposed to improve quality reconstruction, thereby reducing the visibility of compression artifacts on high resolution VDUs. In our technique, the weighting integer values in the intra and inter AQMs are adaptive and contingent upon the resolution of the target VDU. We utilized SHM 9.0 to evaluate the technique on various sequences of different classes. More specifically, we have created two-layered bit-streams, with one BL and one EL, where each layer is aimed at the resolution of a VDU. Compared with anchors, the proposed method yields important coding efficiency and visual quality improvements, with a maximum luma BD-Rate improvement of 56.5% in the EL and a maximum SSIM value difference of 0.8660 vs. 0.8538. In addition, our technique yields modest encoding and decoding time improvements.

## 6. REFERENCES

[1] C. Wang, S. Lee and L. Chang, "Designing JPEG quantization tables based on human visual system," *IEEE International Conference on Image Processing*, Kobe, Japan, 1999, vol. 2, pp. 376-380.

[2] S. Daly, "Subroutine for the generation of a two dimensional human visual contrast sensitivity function," *Technical Report (233203Y), Eastman Kodak*, 1987.

[3] G. Sullivan, J-R. Ohm, W. Han and T. Wiegand, "Overview of the High Efficiency Video Coding (HEVC) Standard," *IEEE Trans. Circuits Syst. Video Technol.*, vol. 22, no. 12, pp. 1649-1668, 2012.

[4] M. Budagavi, A. Fuldseth, G. Bjontegaard, V. Sze and M. Sadafale, "Core Transform Design in the High Efficiency Video Coding (HEVC) Standard," *IEEE J. Sel. Topics Signal Process.*, vol. 7, no. 6, pp. 1649-1668, Dec. 2013.

[5] M. Haque, A. Tabatabai and Y. Morigami, "HVS Model based Default Quantization Matrices," *Document JCTVC-G880*, Geneva, 2011, pp. 1-13.

[6] S. Jeong and B. Jeon, "Newer Quantization Matrices for HEVC," *Document JCTVC-I0126*, Geneva, 2012, pp. 1-8.

[7] M. Li, "Markov Random Field Edge-Centric Image/Video Processing," PhD Thesis, Dept. Electrical Engineering, Univ. California, San Diego., CA, 2007.

[8] N. Casali, M. Naccari, M. Mrak and R. Leonardi, "Adaptive Quantisation in HEVC for Contouring Artefacts Removal in UHD Content," *IEEE International Conference on Image Processing*, Québec, 2015.

[9] V. Sze, M. Budagavi and G. J. Sullivan, "Quantization Matrix," in *High Efficiency Video Coding (HEVC): Algorithms and Architecture*, Springer International Publishing, 2014, pp. 158-159.

[10] J.L. Mannos and D.J. Sakrison, "The effect of a visual fidelity criterion in the encoding of images," *IEEE Trans. Inf. Theory*, vol. 20, pp. 525-536, 1974.

[11] Joint Collaborative Team on Video Coding (2014). SHVC Reference Software, SHM 9.0 [Online]. Available: http://hevc.hhi.fraunhofer.de/

[12] M. Haque and J. A Jackson, "Quantization matrix design for HEVC standard," *European Patent EP2618574A1*, January 8, 2013.

[13] L. Prangnell, V. Sanchez and R. Vanam, "Adaptive Quantization by Soft Thresholding in HEVC," *IEEE Picture Coding Symposium*, Queensland, Australia, 2015, pp. 35-39.

[14] International Telecommunication Union, "Information Technology – Digital Compression And Coding of Continuous-Tone Still Images – Requirements and Guidelines," *CCITT T.81*, September, 1992.

[15] J. Wang, X. Yu and D. He, "On BD-rate calculation," Document *JCTVC-F270*, Torino, 2011.

[16] G. Bjontegaard, "Calculation of Average PSNR Differences Between RD-Curves," *Document VCEG-M33, ITU-T Q.6/SG16 VCEG*, 2001.

[17] G. Bjontegaard, "Improvements of the BD-PSNR model," *Document VCEG-AI11, ITU-T SG16/Q6 VCEG*, 2008.

[18] Z. Wang, A.C. Bovik, H.R. Sheikh, E.P. Simoncelli, "Image quality assessment: from error visibility to structural similarity," *IEEE Trans. Image Processing*, vol. 13, no. 4, pp. 600-612, 2004.

[19] F. Bossen, "Common HM Test Conditions and Software Reference Configurations," *Document JCTVC-L1100*, Geneva, 2013, pp. 1-4.

[20] V. Seregin and Y. He, "Common SHM Test Conditions and Software Reference Configurations," *Document JCTVC-Q1009*, Valencia, 2014, pp. 1-4.

[21] G. J. Sullivan, J. M. Boyce, Ying Chen, J-R. Ohm, C. A. Segall and A. Vetro, "Standardized Extensions of High Efficiency Video Coding (HEVC)," *IEEE J. Sel. Topics Signal Process.*, vol. 7, no. 6, pp. 1001-1016, 2013.

[22] L. Prangnell and V. Sanchez, "Adaptive Quantization Matrices for HD and UHD Display Resolutions in Scalable HEVC," *IEEE Data Compression Conference*, 2016.